\documentclass[journal]{IEEEtran}
\usepackage[T1]{fontenc}% optional T1 font encoding

\usepackage{cite}

\ifCLASSINFOpdf
  \usepackage[pdftex]{graphicx}
  \graphicspath{{./images/}}
\else
\fi

\usepackage{mathtools}
\usepackage{mathrsfs}
\usepackage{siunitx}
\usepackage{bm}
\usepackage{amsbsy}
\usepackage{amsfonts}

\usepackage{algorithm}
\usepackage[noend]{algpseudocode}
\makeatletter
\def\BState{\State\hskip-\ALG@thistlm}
\makeatother

\ifCLASSOPTIONcompsoc
  \usepackage[caption=false,font=normalsize,labelfont=sf,textfont=sf]{subfig}
\else
  \usepackage[caption=false,font=footnotesize]{subfig}
\fi

\usepackage{tabularx}

\usepackage[table,xcdraw]{xcolor}

\usepackage{multirow}

\newcommand{\eexpo}[1]{\exp \left(  #1 \right) }

\newcommand{\elogtwo}[1]{\log_{2} \left(  #1 \right) }

\newcommand{\eprob}[1]{Pr\left(  #1 \right)}

\newcommand{\eave}[2]{\mathbb{E}_{#1} \left\lbrace #2 \right\rbrace}
\newcommand{\eabsn}[2]{\left|  #1 \right| ^{#2}}
\newcommand{\eangle}[1]{\angle\left(  #1 \right)}

\newcommand{\econj}[1]{\left( #1 \right)^{*}}

\newcommand{\egausd}[2]{\mathcal{CN}\left( #1,#2\right) }

\newcommand{\eexpabstwo}[1]{\mathbb{E} \left\lbrace  \left| #1 \right|^2 \right\rbrace}
\newcommand{\eexpabsn}[2]{\mathbb{E} \left\lbrace  \left| #1 \right|^{#2} \right\rbrace}

\newcommand{\ecs}[1]{\left(   #1 \right)  }
\newcommand{\ebs}[1]{\left\lbrace    #1 \right\rbrace   }

\def\mydate{\leavevmode\hbox{\the\year/\twodigits\month/\twodigits\day}}
\def\twodigits#1{\ifnum#1<10 0\fi\the#1}

\newcommand{\eAddFig}[4]{
	\begin{figure}[!t]
		\centering
		\includegraphics[width=#2\linewidth]{#1}
		\vspace{-5mm}
		\caption{#4}
		\label{#3}
	\end{figure}
}

\newcommand{\eAddTwoColFig}[4]{
	\begin{figure*}[!t]
		\centering
		\includegraphics[width=#2\linewidth]{#1}
		\vspace{-6mm}
		\caption{#4}
		\label{#3}
	\end{figure*}
}

\usepackage[acronym]{glossaries}

\newacronym{soa}{SoA}{state of the art}
\newacronym{cep}{CEP}{channel estimation procedure}
\newacronym{ts}{TS}{training sequence}

\newacronym{gps}{GPS}{Global Positioning System}

\newacronym{3gpp}{3GPP}{3rd Generation Partnership Project}
\newacronym{5g}{5G}{Fifth Generation}
\newacronym{6g}{6G}{Sixth Generation}
\newacronym{embb}{eMBB}{enhanced mobile broadband}
\newacronym{urllc}{URLLC}{ultra reliable and low-latency communications}
\newacronym{ritm}{RITM}{RIS-In-The-Middle}
\newacronym{dris}{D-RIS}{defensive RIS}

\newacronym{ssb}{SSB}{synchronization signal block}
\newacronym{pss}{PSS}{primary synchronization signal}
\newacronym{sss}{SSB}{secondary synchronization signal}
\newacronym{pbch}{PBCH}{physical broadcast channel}
\newacronym{prach}{PRACH}{physical random access channel}
\newacronym{dmrs}{DM-RS}{demodulation reference signals}

\newacronym{mmW}{mmWave}{millimetre waves}
\newacronym{cmW}{cmWave}{centimetre waves}
\newacronym{dmW}{dmWave}{decimetre waves}
\newacronym{bmp}{BMP}{beam-management procedure}
\newacronym{iap}{IAP}{initial access procedure}

\newacronym{ris}{RIS}{reconfigurable intelligent surface}
\newacronym{fsm}{FSM}{frequency shifter meta-surface}
\newacronym{comp}{CoMP}{coordinated multi-point}

\newacronym{bs}{BS}{base station}
\newacronym{ue}{UE}{user equipment}
\newacronym{tdl}{TDL}{tapped delay line}
\newacronym{umi}{UMi}{urban micro-cell}

\newacronym{tdm}{TDM}{time-division multiplexing}
\newacronym{fdm}{FDM}{frequency-division multiplexing}
\newacronym{tdd}{TDD}{time-division duplexing}
\newacronym{fdd}{FDD}{frequency-division duplexing}
\newacronym{ul}{UL}{uplink}
\newacronym{dl}{DL}{downlink}
\newacronym{csi}{CSI}{channel state information}

\newacronym{upa}{UPA}{uniform planar array}
\newacronym{ula}{ULA}{uniform linear array}

\newacronym{dft}{DFT}{discrete Fourier transform}
\newacronym{idft}{IDFT}{inverse discrete Fourier transform}

\newacronym{cp}{CP}{cyclic prefix}
\newacronym{ofdm}{OFDM}{orthogonal frequency division multiplexing}

\newacronym{tr}{TR}{tone reservation}

\newacronym{pa}{PA}{power amplifier}
\newacronym{sspa}{SSPA}{solid state power amplifier}

\newacronym{ls}{LS}{Least-Squares}
\newacronym{clt}{CLT}{Central Limit Theorem}

\newacronym{cordic}{CORDIC}{coordinate rotation digital computer}
\newacronym{qcqp}{QCQP}{quadratically constrained quadratic program}

\newacronym{psam}{PSAM}{pilot symbol assisted modulation}
\newacronym{st}{ST}{superimposed training}
\newacronym{htst}{HT-ST}{hollow tone-aided superimposed training}
\newacronym{tlst}{TLST}{two layer superimposed training}

\newacronym{mrt}{MRT}{maximum ratio transmission}
\newacronym{mrc}{MRC}{maximum ratio combining}
\newacronym{zf}{ZF}{zero-forcing}
\newacronym{mimo}{MIMO}{multiple-input multiple-output}

\newacronym{qam}{QAM}{quadrature amplitude modulation}

\newacronym{los}{LOS}{line-of-sigh}
\newacronym{nlos}{NLOS}{non-line-of-sigh}
\newacronym{awgn}{AWGN}{additive white Gaussian noise}
\newacronym{papr}{PAPR}{peak-to-average power ratio}
\newacronym{snr}{SNR}{signal to noise ratio}
\newacronym{sinr}{SINR}{signal to interference and noise ratio}
\newacronym{ber}{BER}{bit error rate}
\newacronym{ser}{SER}{symbol error rate}
\newacronym{mse}{MSE}{mean squared error}
\newacronym{isi}{ISI}{inter-symbol interference}
\newacronym{ici}{ICI}{inter-carrier interference}
\newacronym{asr}{ASR}{achievable secrecy rate}

\begin{document}
\title{ Defensive Reconfigurable Intelligent Surface (D-RIS) Based on Non-Reciprocal Channel Links}
% FRISBEE-6G	FRequency shIfter Surface for uBiquitous cEllular nEtworks in 6G

\author{Kun Chen-Hu,~\IEEEmembership{Member,~IEEE},
	and~Petar~Popovski,~\IEEEmembership{Fellow,~IEEE}% <-this % stops a space
%	\thanks{Copyright (c) 2021 IEEE. Personal use of this material is permitted. However, permission to use this material for any other purposes must be obtained from the IEEE by sending a request to pubs-permissions@ieee.org.}
	\thanks{Kun Chen-Hu and Petar Popovski are with the Department of Electronic Systems, Aalborg University, Denmark. E-mails: \{kchenhu, petarp\}@es.aau.dk.}% <-this % stops a space
	%\thanks{Manuscript received April 19, 2005; revised January 15, 2020.}
	\thanks{This work has been supported by the Villum Investigator Grant “WATER” from the Velux Foundation, Denmark.}
}

\markboth{IEEE Transactions on Communications,~Vol.~XX, No.~XX, July~2024}%
{Shell \MakeLowercase{\textit{et al.}}: Bare Demo of IEEEtran.cls for IEEE Communications Society Journals}

% Make the title area
\maketitle

\begin{abstract}
A reconfigurable intelligent surface (RIS) is commonly made of low-cost passive and reflective meta-materials with excellent beam steering capabilities. It is applied to enhance wireless communication systems as a customizable signal reflector. However, RIS can also be adversely employed to disrupt the existing communication systems by introducing new types of vulnerability to the physical layer. We consider the \emph{RIS-In-The-Middle (RITM) attack}, in which an adversary uses RIS to jeopardize the direct channel between two transceivers by providing an alternative one with higher signal quality. This adversary can eavesdrop on all exchanged data by the legitimate users, but also perform a false data injection to the receiver. This work devises anti-attack techniques based on a non-reciprocal channel produced by a defensive RIS (D-RIS). The proposed precoding and combining methods and the channel estimation procedure for a non-reciprocal link are effective against potential adversaries while keeping the existing advantages of the RIS. We analyse the robustness of the system against attacks in terms of achievable secrecy rate and probability of detecting fake data. We believe that this defensive role of RIS can be a basis for new protocols and algorithms in the area.\end{abstract}

\begin{IEEEkeywords}
channel estimation, defensive, meta-surface, non-reciprocity, RIS.
\end{IEEEkeywords}

% For peer-reviewed papers, you can put extra information on the cover
% \fi
%
% For peer review papers, this IEEEtran command inserts a page break and
\IEEEpeerreviewmaketitle

\section{Introduction}

\IEEEPARstart{R}{econfigurable} intelligent surface (RIS) \cite{ris1,ris2,ris3,ris4,ris5,ris6,ris7,riskey} is a structure made of low-cost passive meta-materials that exhibit configurable electromagnetic properties. These tunable passive elements can exhibit excellent beam steering capabilities and extremely pencil-shape narrow beams. RIS has been extensively applied to enhance wireless communication systems, as it can transform traditional wireless networks into smart radio environments by manipulating the propagation environment. It can enlarge the coverage distance of the signal by providing an alternative propagation path with a better channel condition and/or enhance the data rate of the existing links. This is especially interesting in millimetre and terahertz bands since blockage is very frequent due to high attenuation. Due to this, RIS has been seen as an important element in the evolution to \acrfull{6g} \cite{6gsec,6gind,6gvr}. In addition, RIS surfaces are cost-competitive as compared to the traditional relays, enabling massive deployment of this technology. These reflective panels are easy to integrate with the aesthetics of the environment since they can take the shape of wallpapers, window glasses, building facades and roadside billboards. 

On the flip side, low-cost passive elements can be easily camouflaged in the environment, such that the RIS technology can also be adversely employed to disrupt the existing communication systems by introducing new types of vulnerability to the physical layer. \emph{\acrfull{ritm} attack} \cite{ristest1,ristest2,riseve1,riseve2,riseve3,riseve4,riseve5} consists of using a metasurface to divert and usurp an existing established communication link between two legitimate transceivers thanks to its advanced control and manipulation of electromagnetic waves at the physical layer. 
Several works in the literature have assumed that the employed cryptographic system is typically not robust enough against the adversary, since strong encryption methods have an unaffordable complexity.
It has been shown experimentally \cite{ristest1,ristest2} that an adversarial RIS panel is not only able to eavesdrop on all the information exchanged by these legitimate transceivers, but it is also performing a false data injection, by reflecting either a corrupted or faked version of the received information. Moreover, no traces are left by the adversarial panel, making it impossible to be detected by the traditional existing procedures in the physical layer.

As a solution to partially alleviate these security issues, the deployment of \acrfull{dris} is used to produce an alternative propagation path. If the size of the D-RIS is significantly larger than the adversarial RIS, the signal manipulated by the adversary is eclipsed by the legitimate one. The theoretical analysis performance of this solution \cite{riseve1,riseve2} is given in terms of \acrfull{asr}. However, they cannot guarantee that the size of the legitimate RIS is going to be always larger and prevent data eavesdropping. Other works targeted the prevention of eavesdroppers by using narrow directive beams in the transmitter and D-RIS \cite{riseve3,riseve4,riseve5}, and hence spatially filter the adversary. The joint optimization problem to compute these precoders is typically assuming the availability of the instantaneous \cite{riseve3} or statistical \cite{riseve4} \acrfull{csi} of the adversary, or the geographical localization of the adversary \cite{riseve5}. These may be hard to justify as adversaries can be easily camouflaged in the environment. Later, experiments showed that a mmWave system could be easily disrupted by the RITM attack \cite{ristest2}, even using narrow beam widths. Moreover, in the hypothetical case that eavesdropping was solved, the adversary can still perform a false data injection to trick the receiver, and this issue, to the best of our knowledge, has not been addressed yet. Consequently, these new potential threats produced by the environmental manipulation at the RIS need to be carefully explored and analysed.

To the best of our knowledge, these security and privacy issues related to eavesdropping and false data injection by considering an unknown adversary have not been thoroughly addressed in previous research works. Therefore, the main objective of this paper is to develop novel defensive techniques in the physical layer, relying on the non-reciprocal channels produced by a D-RIS in the absence of any information related to the adversary. Note that the concept of exploiting a non-reciprocal channel for communications has never been proposed for communications before. Since the direct channel between the legitimate devices is reciprocal, the adversary will replicate an alternative reciprocal one with better channel propagation conditions. The non-reciprocal channel response may be obtained in real-time by properly configuring the phase configurations, and the CSI of the \acrfull{dl} and \acrfull{ul} can be seen as a pair of keys for each D-RIS link. To exploit the non-reciprocity, both legitimate entities must properly precode the data stream to be transmitted and combine the received symbols using their unique combination of the CSI of both links. Hence, any illegitimate intruder enhanced by an adversarial RIS cannot replicate this non-reciprocal channel since it does not have the secret CSI. The receiver can trust that the received information is sent by a legitimate entity, as it can filter out illegitimate transmissions and detect potential adversaries by using the CSI of the non-reciprocal channel.

The main contributions are summarized as follows:
\begin{itemize}
	
	\item The proposal of a defensive system against RITM attacks based on a novel concept of a non-reciprocal channel provided by a D-RIS is presented. Both legitimate entities can exploit this non-reciprocal property of the channel to avoid eavesdropping and fake data injection performed by any adversary since the CSI of both DL and UL for each UE of the cell are unique. Additionally, an adversary requires more resources to jeopardize a link as a consequence of jointly exploiting the CSI of both DL and UL.
	
	\item Given the non-reciprocal channel between the BS and UE, it is proposed that the precoding technique is a variant of the \acrfull{mrt} \cite{mimo_mrt}, based on the joint exploitation of the CSI of both links. Then, not only the channel effects of the received signal has been removed due to the precoding at the receiver, but the proposed combiner is also checking its integrity by using the unique CSI of each entity. Consequently, the CSI of the channels plays the role of a secret key, while precoding and combining emulate the encryption and decryption processes at the physical layer.
	
	\item A novel \acrfull{cep} for a non-reciprocal channel via D-RIS is presented. 
	% Since the CSI of the DL and UL channels should be secret, CSI feedback is no longer allowed to avoid its interception. 
	To allow the CSI acquisition of DL and UL at the BS and UE, phase flipping at D-RIS is proposed. This method consists of exchanging the phase values of the D-RIS configured for the DL and UL to the UL and DL, respectively. This proposal is not only safer as compared to the traditional CSI feedback, but it is also more efficient since the transmission of the estimated CSI requires more resources than sending a training sequence.
	
	\item An analysis of the system performance is provided to show the benefits of the proposed technique as compared to the traditional one based on a reciprocal channel. The ASR~\cite{sec} is evaluated to address the eavesdropping. Additionally, the probability of decoding false symbols, injected by the adversarial RIS, is also analysed to characterize the trickery capabilities of the adversary. The performance is enhanced by exploiting the non-reciprocal channel produced by the D-RIS, and the precoding/combining techniques can effectively avoid eavesdropping and data manipulation as it augments the time required to find out the precoders/combiners.
	
	\item Finally, a performance assessment based on numerical evaluation is obtained in terms of ASR and probability of detecting injected false symbols. The results not only verify the theoretical analysis but also highlight that the proposed architecture significantly outperforms the existing technique in the literature, making it a reliable system to be used in future wireless networks.
\end{itemize}

The remainder of the paper is organized as follows. 
Section \ref{sec:sys} introduces the system model of the considered D-RIS in a RITM attack scenario. 
Section \ref{sec:attack} details the operations carried out by a potential adversary. 
Sections \ref{sec:anti} and \ref{sec:chan} provide the novel defensive technique and the CEP, respectively, based on a non-reciprocal link via D-RIS.
Section \ref{sec:sar} analyses the system performance of the proposed system in terms of ASR and the probability of detecting fake symbols. 
Section \ref{sec:per} presents several numerical results for the proposed architecture, providing an assessment of the achieved performance. 
Finally, in Section \ref{sec:conclusion}, the conclusions are disclosed.

\textbf{Notation:} scalar quantities are denoted by normal letters.
%$\left[ \mathbf{A}\right]_{m,n}$ denotes the element in the $m$-th row and $n$-th column of $\mathbf{A}$.
%$\left[ \mathbf{a}\right]_{n}$ represents the $n$-th element of vector $\mathbf{a}$.
%%$\left[ \mathbf{A}\right]_{m,\bullet}$ and $\left[ \mathbf{A}\right]_{\bullet,n}$ denote the $m$-th row vector and $n$-th column vector of $\mathbf{A}$, respectively.
%$\mathbf{I}_{M}$ is the identity matrix of size $\esize{M}{M}$. 
%%$\mathbf{0}_{M,N}$ is the zero matrix of size  $\esize{M}{N}$.
%$\eones{M}{N}$ denotes a matrix of ones of size $\esize{M}{N}$.
%$\mathbf{A}=\ediag{\mathbf{a}}$ is a diagonal matrix whose diagonal elements are formed by the elements of vector $\mathbf{a}$.
%%$\edet{\cdotp}$ represents the determinant.
%$\etra{\cdotp}$ corresponds to the matrix trace operation.
%The superscripts $\etras{\cdotp}$ and $\eherm{\cdotp}$ denote transpose and hermitian operations, respectively.
The superscript $\econj{\cdotp}$ denotes a complex conjugate operation.
%%The superscripts $\etras{\cdotp}$,$\eherm{\cdotp}$, $\econj{\cdotp}$ and $\left( \cdotp \right) ^{\dagger}$ denote transpose, hermitian, complex conjugate and Moore-Penrose inverse operations, respectively.
%% $\Re(\cdotp)$ and $\Im(\cdotp)$ represent the real and imaginary part, respectively.
%$*$ denotes the convolution operation. 
%$\circledast$ is the circular convolution operation. 
%$\otimes$ corresponds to the Kronecker product of two matrices.
%%$\circ$ is the Hadamard product of two matrices.
$\eave{}{\cdotp}$ represents the expected value.
%%$\text{Var}\left\lbrace \cdotp \right\rbrace$ denotes variance.
$\mathcal{CN}(0,\sigma^2)$ corresponds the circularly-symmetric and zero-mean complex normal distribution with variance $\sigma^2$. 
%$\enormn{\cdotp}{}{p}$ denotes the $p$-norm.
%$\eabsn{\cdot}{}$ is the absolute value of a complex number.
$\eabsn{\cdot}{}$ and $\eangle{\cdot}$ represent the absolute value and phase of a complex number, respectively.
%%$\eceil{x}$ represents the smallest integer greater than or equal to $x$.
%%$\efloor{x}$ represents the largest integer less than or equal to $x$.
%$\emod{m}{n}$ is the integer modulo operator which provides the remainder of the ratio $m/n$.
%%$\emax{\mathbf{a}}$ corresponds to the maximum value of the vector $\mathbf{a}$.
%$\mathbb{C}^{K}$ and $\mathbb{R}^{K}$ are $K$-dimensional complex and real spaces, respectively, and $\mathbb{C}^{K\times K}$ is the $K\times K$-dimensional complex space.

\section{System Model of the Proposed Defensive Scenario under RITM Attack }
\label{sec:sys}

A BS is serving a particular fixed UE of interest, and it is assumed that the UE has successfully authenticated into the system. Both legitimate entities are equipped with a single omnidirectional antenna, whose pattern corresponds to a pessimistic case as compared to \cite{ristest1,ristest2,riseve1,riseve2,riseve3,riseve4,riseve5} since any adversary can attack the legitimate communication link from any geographical position. A D-RIS is deployed to not only improve the quality of the channel between BS and UE by providing a new alternative one, but it also enhances the security by neutralizing potential adversaries, see Fig. \ref{fig:system}. The D-RIS is equipped by a \acrfull{upa}, whose number of elements is given by $M_{a} = M_{a}^{x}M_{a}^{y}$, where $M_{a}^{x}$ and $M_{a}^{y}$ are the number of elements in the $x$ and $y$-axis, respectively. In addition to the legitimate entities (BS, UE and D-RIS), there is an adversary Eve, who is interested in jeopardizing the legitimate link between the BS and UE. According to \cite{ristest1}, Eve is also equipped with another adversarial RIS, which is capable of manipulating the impinging signal and reflecting an alternative fake one of an already established communication link. For simplicity, it is assumed that the hardware specifications related to the adversarial RIS of Eve are similar to the D-RIS, where the number of elements of the UPA is denoted by $M_{e}$. In this work, any information related to Eve is considered to be unknown by legitimate entities, such as geographical positions, RIS size, CSI, etc., unlike \cite{riseve3,riseve4,riseve5}, which is unrealistic.

The chosen waveform corresponds to the well-known \acrfull{ofdm}, which is deployed in the current 5G \cite{nr-211}. It is considered that a slot is the minimum allocable resource in the system, which is built by $K$ subcarriers and $N$ consecutive OFDM symbols. A \acrfull{tdd} scheme is adopted \cite{mimo_tdd}, where the DL and UL transmissions are allocated in different OFDM symbols within one slot. Since all involved entities (BS, UE, D-RIS, and Eve) are fixed in a specific position within a delimited area, it is assumed that the coherence time of all reciprocal channels is long enough to cope within, at least, one slot. Moreover, it is considered that the \acrfull{cp} is long enough to absorb the multi-path effects of all channels, avoiding the \acrfull{ici} and \acrfull{isi} in the OFDM signal. Consequently, without loss of generality and for the sake of the space, a particular subcarrier of the OFDM symbol, out of $K$, is taken into consideration.

\eAddFig{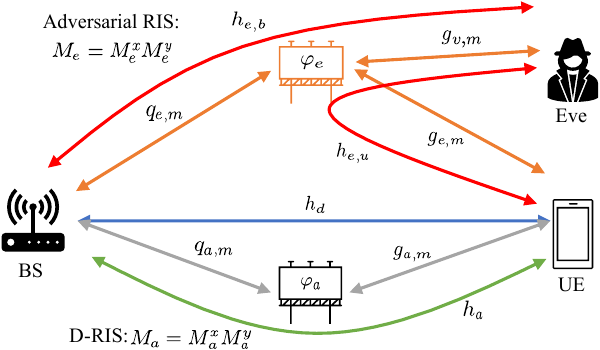}{1}{fig:system}{The system model is built by a legitimate link between BS and UE assisted by a D-RIS. Then, Eve is a third illegitimate entity, that is interested in jeopardizing the legitimate link.}

At the receiver, after discarding the CP and performing the \acrfull{dft}, the received signal of the UE and BS at the $n$-th OFDM symbol and the subcarrier of interest can be modelled as
\begin{equation} \label{eq:yudl}
	y_{u,n} = h_{r,n} x_{b,n} + w_{u,n}, \quad n \in \mathcal{N}^{\text{DL}},
\end{equation}
\begin{equation} \label{eq:ybul}
	y_{b,n} = h_{r,n} x_{u,n} + w_{b,n}, \quad n \in \mathcal{N}^{\text{UL}},
\end{equation}
respectively, where $x_{b,n}$ and $x_{u,n}$ are the transmitted data symbols from the BS and UE at $n$-th OFDM symbol and the subcarrier of interest, respectively. $w_{u,n}$ and $w_{b,n}$ account for the \acrfull{awgn} of the UE and BS at $n$-th OFDM symbol and the subcarrier of interest, respectively, whose distribution follows $\egausd{0}{\sigma_{w}^{2}}$. $\mathcal{N}^{\text{DL}}$ and $\mathcal{N}^{\text{UL}}$ are the two subsets that contain the OFDM symbol indices for the DL and UL transmissions within one slot, respectively. $h_{r,n}$ is the resulting channel frequency response between BS$\leftrightarrow$UE at $n$-th OFDM symbol and the subcarrier of interest, which can be expressed as
\begin{equation} \label{eq:hr}
	h_{r,n} = h_{d} + h_{a,n} + h_{e,n}, \quad n \in \mathcal{N},
\end{equation}
\begin{equation} \label{eq:setn}
	\mathcal{N} = \ebs{0, 1, \cdots, N} = \mathcal{N}^{\text{DL}} \cup \mathcal{N}^{\text{UL}}, \quad \mathcal{N}^{\text{DL}} \cap \mathcal{N}^{\text{UL}} = \emptyset,
\end{equation}
where $\mathcal{N}$ is the full set that contains all the $N$ OFDM symbol indices within one slot. $h_{d}$ corresponds to the frequency channel response of the direct link between the BS$\leftrightarrow$UE at the subcarrier of interest, whose distribution follows $\egausd{0}{\sigma_{d}^{2}}$. 
$h_{a,n}$ is the alternative cascaded channel frequency response \cite{ris3} between the legitimate entities via D-RIS (BS$\leftrightarrow$D-RIS$\leftrightarrow$UE) at the $n$-th OFDM symbol and the subcarrier of interest, which can be modelled as
\begin{equation} \label{eq:ha}
	h_{a,n} =  \sum_{m=1}^{M_{a}} \eexpo{j\varphi_{a,m,n}} q_{a,m} g_{a,m}, \quad n\in \mathcal{N},
\end{equation}
\begin{equation} \label{eq:ha_distri}
q_{a,m} \sim \egausd{0}{\sigma_{qa}^{2}}, \quad  g_{a,m} \sim \egausd{0}{\sigma_{ga}^{2}},
\end{equation}
where $\varphi_{a,m,n}$ is the tunable phase response of the $m$-th element of the D-RIS at the $n$-th OFDM symbol, $q_{a,m}$ and $g_{a,m}$ are the reciprocal channel frequency response of BS$\leftrightarrow$D-RIS and D-RIS$\leftrightarrow$UE at $n$-th OFDM symbol and the interested subcarrier, respectively. Additionally, the channel coefficients among different passive elements of the RIS are considered to be independently and identically distributed to simplify the notation. Analogously to D-RIS, $h_{e,n}$ represents the cascaded frequency channel response between each legitimate entity and Eve via adversarial RIS, which can be modelled by as
\begin{equation} \label{eq:he}
	h_{e,n} = \begin{cases} 
		h_{e,u,n} & n \in \mathcal{N}^{\text{DL}} \\ 
		h_{e,b,n} & n \in \mathcal{N}^{\text{UL}}
	\end{cases},
\end{equation}
\begin{equation} \label{eq:hed}
	h_{e,u,n} = \sum_{m=1}^{M_{a}} \eexpo{j\varphi_{e,m,n}} g_{v,m} g_{e,m},
\end{equation}
\begin{equation} \label{eq:heu}
	h_{e,b,n} =  \sum_{m=1}^{M_{a}} \eexpo{j\varphi_{e,m,n}} g_{v,m} q_{e,m},
\end{equation}
\begin{equation} \label{eq:he_distri}
	g_{v,m} \sim \egausd{0}{\sigma_{gv}^{2}},
\end{equation}
where $\varphi_{e,m,n}$ is the tunable phase response of the $m$-th element of the D-RIS at the $n$-th OFDM symbol, $h_{e,u,n}$ and $h_{e,b,n}$ are the reciprocal channel frequency responses of UE$\leftrightarrow$Eve and BS$\leftrightarrow$Eve at the at $n$-th OFDM symbol and the interested subcarrier, respectively. $g_{v,m}$ denotes the channel frequency response between the adversarial RIS and Eve. Both RISs require a training period \cite{ris4,ris5} to find out their respective best phase configurations ($\varphi_{a,m,n}$ and $\varphi_{e,m,n}$), and ensure that their resulting channel gains are much higher than the direct one ($\eabsn{h_{a,n}}{}, \eabsn{h_{e,n}}{} >> \eabsn{h_{d}}{}$). 

Note that it is assumed that the metasurfaces equipped by both D-RIS and Eve are fully passive, and hence only the phases of these passive elements can be manipulated. The reason behind avoiding the additional amplitude manipulation is due to the energy and complexity issues, among others. On one side, the amplification of the impinging signal requires extra power consumption and also increases the cost of the panel. On the other hand, finding the best phase configuration becomes more complex, while the training period will be also significantly increased within the channel coherence time \cite{rappa}. This reduces the effective operational time of the panel, such that there is always a tradeoff between the performance of RIS and the required training time to obtain the best amplitude and phase configurations.

The channel between the legitimate entities via D-RIS ($h_{a,n}$) can be turned into a non-reciprocal one thanks to the proper configuration of the tunable phase configurations given in (\ref{eq:ha}), which can be decomposed as
\begin{equation} \label{eq:tvphase1}
	\varphi_{a,m,n} = \varphi_{a,m} + \varphi_{a,n}, \quad 1 \leq m \leq M_{a}, \quad 1 \leq n \leq N, 
\end{equation}
where $\varphi_{a,m}$ is the static phase at the $m$-th passive element, which is typically designed to create an alternative propagation path between the BS and UE \cite{ris1,ris2,ris3,ris4,ris5,ris6,ris7,riskey}, meanwhile $\varphi_{a,n}$ corresponds to the common dynamic phase which is the same for all passive elements of the D-RIS. Therefore, (\ref{eq:ha}) can be rewritten as
\begin{equation} \label{eq:ha2}
	\begin{split}
		h_{a,n} & = \eexpo{j\varphi_{a,n}} \sum_{m=1}^{M_{a}} \eexpo{j\varphi_{a,m}} q_{a,m} g_{a,m} \\
		& = \eexpo{j\varphi_{a,n}} h_{a}, \quad n\in \mathcal{N},
	\end{split}
\end{equation}
where $h_{a}$ is the traditional reciprocal cascaded channel frequency response between the legitimate entities via D-RIS (BS$\leftrightarrow$D-RIS$\leftrightarrow$UE) at a subcarrier of interest, whose gain can be maximized by tuning each phase configuration of the D-RIS ($\varphi_{a,m}$) \cite{ris1,ris2,ris3,ris4,ris5,ris6,ris7,riskey}, meanwhile the additional factor $\eexpo{j\varphi_{a,n}}$ is producing the non-reciprocal effect in the channel response. This time-varying phase at the D-RIS can be tuned by the scheduler at the BS, where its value is changed according to the DL and UL OFDM symbol periods as
\begin{equation} \label{eq:tvphase2}
	\varphi_{a,n} = \begin{cases} 
		\varphi_{a}^{\text{DL}} & n \in \mathcal{N}^{\text{DL}} \\ 
		\varphi_{a}^{\text{UL}} & n \in \mathcal{N}^{\text{UL}}
	\end{cases} \rightarrow 
	h_{a,n} = \begin{cases} 
		h_{a}^{\text{DL}} & n \in \mathcal{N}^{\text{DL}} \\ 
		h_{a}^{\text{UL}} & n \in \mathcal{N}^{\text{UL}}
	\end{cases},
\end{equation}
where $\varphi_{a}^{\text{DL}}, \varphi_{a}^{\text{UL}}$ are the common time-varying phase configurations for the DL and UL, respectively, and $h_{a}^{\text{DL}}, h_{a}^{\text{UL}}$ account for the cascaded channel frequency response between the legitimate entities via D-RIS at the subcarrier of interested for the DL and UL, respectively. This non-reciprocal channel created by the D-RIS will be exploited to provide a reliable network between legitimate entities since it allows the detection of potential adversaries and the avoidance of eavesdropping and data manipulation of exchanged information between BS and UE.

\section{RITM Attack: Eavesdropping and Data Manipulation}
\label{sec:attack}

\begin{figure}[!t]
	\centering
	\subfloat[RITM attack in DL. The BS transmits a precoded message.]
	{\includegraphics[width=1\linewidth]{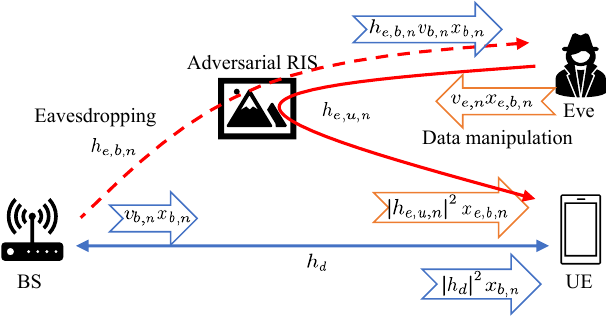}%
		\label{fig:introDownAttack}}
	\hfil
	\subfloat[RITM attack in UL. The UE directly transmits the message.]
	{\includegraphics[width=1\linewidth]{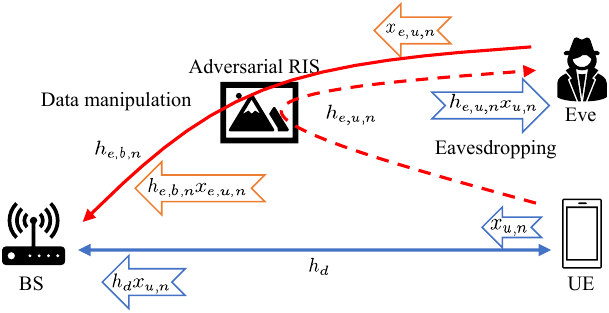}%
		\label{fig:introUpAttack}}
	\caption{The adversary Eve is the illegitimate entity that is listening and manipulating the data exchanged between the BS and UE.}
	\label{fig:introAttack}
\end{figure}

Given an already established legitimate communication link between a BS and a particular UE of interest, an adversary (Eve) is interested in jeopardizing this link (see Fig. \ref{fig:introAttack}). It is considered that Eve is equipped with an adversarial RIS, which can be easily camouflaged in the environment \cite{ristest1,ristest2,riseve1,riseve2,riseve3,riseve4,riseve5}. First, Eve is capable of eavesdropping the information transmitted by legitimate entities, disrupting the privacy of the communication. Then, once Eve has collected enough information by listening to both control and data channels, she is also able to replace the original symbols by transmitting a new manipulated stream to both legitimate entities, leading to data pollution.

In a TDD scheme, channel reciprocity can be assumed. The CSI is typically estimated in the UL and reused in the DL. Initially, the BS applies the well-known \acrfull{ls} \cite{leastsquares} to the received pilot symbols to obtain the CSI. Then, it is also responsible for performing precoding and combining techniques to its transmitted and received symbols, respectively. The UE is released of these procedures to keep its complexity as low as possible and enlarge its autonomy. Taking into account this operation, the potential attacking operations performed by Eve are described in the following two subsections.

\subsection{CSI Acquisition}
Before jeopardizing the legitimate link, Eve is obtaining the CSI of the links between BS$\leftrightarrow$Eve ($\widehat{h}_{e,b,n}$) and UE$\leftrightarrow$Eve ($\widehat{h}_{e,b,n}$) by listening to the reference signals transmitted by the legitimate entities and performing an LS \cite{leastsquares} at the receiver. Then, Eve intercepts the precoded symbols sent by the BS. Assuming that Eve has a high-performance CPU to execute some advanced optimization algorithms, Eve is capable of retrieving the chosen precoding matrix by the BS and its corresponding reciprocal channel between BS$\leftrightarrow$UE ($\widehat{h}_{d}$). Alternatively, Eve can also obtain the CSI of the legitimate link by directly intercepting those symbols that contain the CSI feedback sent by the UE. For the sake of simplicity and ease of the notation, it is assumed that all CSI obtained by Eve are perfectly estimated. Hence, the transmission of CSI feedback should be forbidden since it is actively helping Eve to jeopardize the legitimate link, and an alternative CEP is presented in Section \ref{sec:chan}.

\subsection{DL: Precoding at the BS}

In the case of DL, the complex symbols transmitted by the BS ($x_{b,n}$), which belong to a \acrfull{qam} constellation, must be precoded. Later, the UE can directly make a symbol decision without performing any additional operation since the channel is compensated by the precoder applied at the BS (see Fig. \ref{fig:introDownAttack}). Therefore, the received symbols at Eve in the DL ($y_{e,n}$) can be expressed as
\begin{equation} \label{eq:eve_dl1}
	y_{e,n} = h_{e,b,n} v_{b,n} x_{b,n} + w_{e,n}, \quad n \in \mathcal{N}^{\text{DL}}.
\end{equation}
where $v_{b,n}$ is the chosen precoder at $n$-th OFDM symbol and the particular subcarrier of interest in the BS and $w_{e,n}$ account for the \acrfull{awgn} of Eve at $n$-th OFDM symbol and the subcarrier of interest, whose distribution follows $\egausd{0}{\sigma_{w}^{2}}$. Consequently, Eve can retrieve the sent symbols by the BS if the precoder matrix is exposed. As a solution, robust enough precoding and combining methods based on exploiting the non-reciprocal channel via D-RIS are presented in Section \ref{sec:anti}, which are capable of significantly enlarging the time that Eve requires to compute these precoders/combiners, even larger than the channel coherence time.

Moreover, Eve is injecting false data symbols via the adversarial RIS, and hence, the received signal by the UE is given by
\begin{equation} \label{eq:eve_dl2}
	y_{u,n} =  \eabsn{h_{d}}{2}x_{b,n} + h_{e,u,n} v_{e,n}x_{e,b,n} + w_{u,n}, \quad n \in \mathcal{N}^{\text{DL}},
\end{equation}
where $x_{e,b,n}$ denotes the manipulated data stream sent by Eve to replace the data sent by the BS, $v_{e,n}$ is the precoding matrix designed by Eve, and assuming again the MRT criterion \cite{mimo_mrt}, its expression is $v_{e,n} = \econj{h_{e,u,n}}$. Therefore, (\ref{eq:eve_dl2}) can be simplified as
\begin{equation} \label{eq:eve_dl3}
	y_{u,n} =  \eabsn{h_{d}}{2}x_{b,n} + \eabsn{h_{e,u,n}}{2}x_{e,b,n} + w_{u,n}, \quad n \in \mathcal{N}^{\text{DL}}.
\end{equation}

Inspecting (\ref{eq:eve_dl3}), if the signal strength sent by Eve is higher than the original one ($\eabsn{h_{e,u,n}}{} >> \eabsn{h_{d}}{}$), the UE will decode the fake one. Otherwise, the fake signal is behaving as an additional interference. Note that even though Eve is equipped with a passive RIS, its fake signal may have a higher gain than the direct one. This fact is due to the that the direct channel may correspond to a \acrfull{nlos} due to the potential obstacles and the adversarial RIS could be large ($M_{e}\uparrow$) and it can provide a very narrow directive beam pointing to both legitimate entities \cite{ris1}.

\subsection{UL: Combining at the BS}
In this UL scenario, the UE is directly sending the complex symbols ($x_{u,n}$) which also belong to a QAM constellation. Then, the BS combines the received symbol before making the symbol decision (see Fig. \ref{fig:introUpAttack}). Therefore, the eavesdropped signal by Eve can be modelled as
\begin{equation} \label{eq:eve_ul1}
	y_{e,n} = h_{e,u,n} x_{u,n} + w_{e,n}, \quad n \in \mathcal{N}^{\text{UL}},
\end{equation}
where it points out again that Eve can easily retrieve the symbols sent by the UE, and privacy is seriously exposed. 

Then, Eve is also injecting false data into the BS. Analogously to the DL case, the received signal at the BS, given in (\ref{eq:ybul}), can be rewritten as
\begin{equation} \label{eq:eve_ul2}
	y_{b,n} = h_{d} x_{u,n} + h_{e,b,n} x_{e,u,n} + w_{b,n}, \quad n \in \mathcal{N}^{\text{UL}},
\end{equation}
where $x_{e,u,n}$ denotes the manipulated data stream by Eve to replace the data sent by the UE. Again, two streams are arriving at the BS, both the original one transmitted by the UE and the false one injected by Eve. The BS will combine the received signal with the combiner as
\begin{equation} \label{eq:eve_ul4}
	z_{b,n} = v_{b,n}y_{b,n} \quad n \in \mathcal{N}^{\text{UL}},
\end{equation}
where $v_{b,n}$ is the combiner chosen at the BS to mitigate the effect of the channel. Again, depending on the signal strength of the direct and channel of Eve, the performance of the system may be seriously compromised if the BS is computing the combiner by using the channel of Eve ($v_{b,n}=\econj{h_{e,b,n}}$), and hence, the BS will demodulate the fake symbols injected by Eve, instead, those transmitted by the UE.

\section{Proposed Defensive System against RITM Attack based on Non-Reciprocal Channel via RIS}
\label{sec:anti}

\eAddTwoColFig{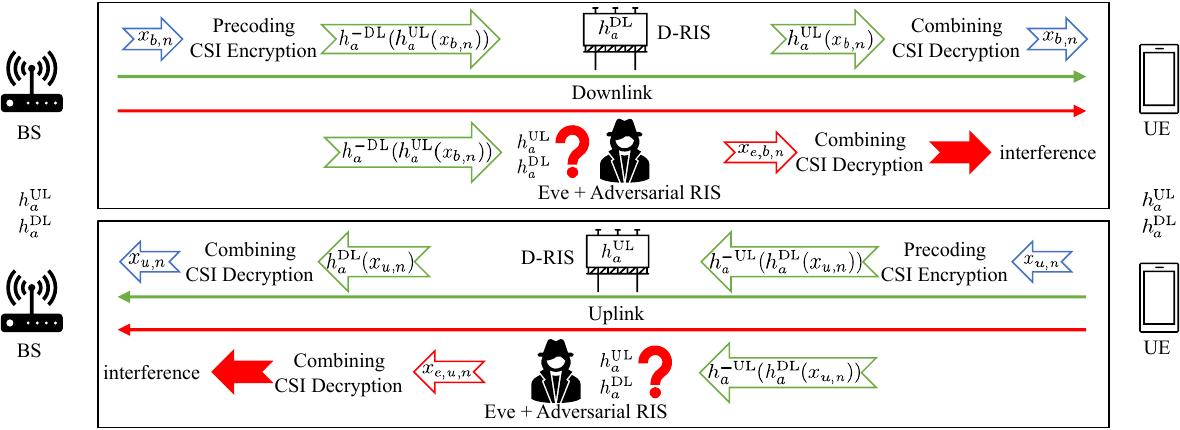}{1}{fig:antiAttack}{Proposed defensive system against RITM attack based on a non-reciprocal channel via RIS. The precoding and combining techniques are based on the CSI of DL and UL, which are produced by the non-reciprocal phase configurations of the D-RIS.}

A novel defensive system against RITM attack relying on a non-reciprocal channel via RIS is depicted in Fig. \ref{fig:antiAttack}, which is capable of solving all the relevant security and privacy issues in the physical layer, described in Section \ref{sec:attack}. This non-reciprocity property makes the D-RIS path completely different from the existing reciprocal ones (direct and Eve's path), and hence, this difference will ensure that legitimate entities will exchange their information through the desired D-RIS link, providing a reliable physical link to the higher layers.

The proposed defensive system assumes that both BS and UE have CSI of both DL and UL channels via D-RIS. At the transmitter, the legitimate entities are performing a precoding (\ref{eq:eve_dl1}) by combining the CSI of these two links. At the receiver, the effect of the precoder is partially mitigated by the channel itself, and the remaining part should be compensated by the combiner at the receiver (\ref{eq:eve_ul4}). The advantage of this procedure is twofold, on one hand, the precoder is more robust than (\ref{eq:eve_dl1}) since Eve requires even more time to find out the CSI of both links, making the eavesdropping process significantly harder. On the other hand, the combining (\ref{eq:eve_ul4}) checks the integrity of the received symbols and verifies that they are coming from the expected entity since the CSI of each UE is unique. In the case that Eve was injecting a false sequence, the combiner would mask it and change the content of these false symbols, and therefore, the decision over these fake data would be wrong. Consequently, the CSI of the DL and UL play the role of secret keys while the precoding and combining procedures are equivalent to encryption and decryption methods in the physical layer, respectively.

For clarity and easier notation, it is assumed that there exists a CEP capable of retrieving the CSI for all involved links considered in Section \ref{sec:sys}, whose details are given in the following Section \ref{sec:chan}. During the CSI acquisition process, two possible scenarios can be considered due to the presence of Eve, namely interference-free and polluted CSI of the D-RIS link. The former corresponds to the regular channel estimation case, where Eve is not able to interfere in the CEP, while the latter fails in the CEPs due to Eve. The proposed defensive method against RITM attack is described in the following two subsections according to the considered scenarios. 

\subsection{Interference-free CSI of the D-RIS Link}
\label{sec:anti1}
In this case, both BS and UE could satisfactorily estimate the CSI of the D-RIS channel ($h_{a}^{\text{DL}}, h_{a}^{\text{UL}}$). Note that it is assumed that the CSI is perfectly obtained to ease the notation. The precoders at the BS ($v_{b,n}$) and UE ($v_{u,n}$) can be computed as
\begin{equation} \label{eq:v_dl_i1}
	v_{b,n} = v_{b,n}^{\text{DL}} v_{b,n}^{\text{UL}} = \econj{h_{a}^{\text{DL}}}\eexpo{j\theta_{a}^{\text{UL}}}, \quad n \in \mathcal{N}^{\text{DL}},
\end{equation}
\begin{equation} \label{eq:v_ul_i1}
	v_{u,n} = v_{u,n}^{\text{UL}} v_{u,n}^{\text{DL}} = \econj{h_{a}^{\text{UL}}}\eexpo{j\theta_{a}^{\text{DL}}}, \quad n \in \mathcal{N}^{\text{UL}},
\end{equation}
\begin{equation} \label{eq:v_i3}
	\theta_{a}^{\text{UL}} = \eangle{h_{a}^{\text{UL}}}, \quad \theta_{a}^{\text{DL}} = \eangle{h_{a}^{\text{DL}}},
\end{equation}
respectively, where $v_{b,n}^{\text{DL}}$ and $v_{b,n}^{\text{UL}}$ are the precoders at the BS obtained from the CSI of DL and UL, respectively. $v_{u,n}^{\text{UL}}$ and $v_{u,n}^{\text{DL}}$ correspond to the precoders at the UE obtained from the CSI of UL and DL, respectively. $\theta_{a}^{\text{DL}}$ and  $\theta_{a}^{\text{UL}}$ account for the phase components of the D-RIS channel in the DL and UL, respectively. Note that (\ref{eq:v_dl_i1})-(\ref{eq:v_i3}) correspond to a modified version of the well-known MRT precoding \cite{mimo_mrt}, where an additional phase ($\theta_{a}^{\text{UL}}, \theta_{a}^{\text{DL}}$) is embedded in the precoders of the BS and UE ($v_{b,n}$ and $v_{u,n}$, respectively).

Taking into consideration (\ref{eq:v_dl_i1})-(\ref{eq:v_i3}) and making use of (\ref{eq:eve_dl3}), the received symbols at the UE and BS, given in (\ref{eq:yudl}) and (\ref{eq:ybul}) respectively, can be rewritten as
\begin{equation} \label{eq:dl_freq_i1}
	y_{u,n} = \eabsn{h_{a}^{\text{DL}}}{2} v_{b,n}^{\text{UL}} x_{b,n} + \eabsn{h_{e,n}}{2} x_{e,b,n} + w_{u}, \quad n \in \mathcal{N}^{\text{DL}},
\end{equation}
\begin{equation} \label{eq:ul_freq_i1}
	y_{b,n} = \eabsn{h_{a}^{\text{UL}}}{2} v_{u,n}^{\text{DL}} x_{u,n} + \eabsn{h_{e,n}}{2} x_{e,u,n} + w_{b}, \quad n \in \mathcal{N}^{\text{UL}},
\end{equation}
respectively. Before performing the symbol decision, a combining based on phase rotation is performed to the received signals at the UE and BS as
\begin{equation} \label{eq:dl_freq_i2}
	\begin{split}
		z_{u,n} & = \eexpo{-j\theta_{a}^{\text{UL}} } y_{u,n} = \eabsn{h_{a}^{\text{DL}}}{2} x_{b,n} \\
		& + \eexpo{-j\theta_{a}^{\text{UL}}} \ecs{\eabsn{h_{e,n}}{2} x_{e,b,n} + w_{u,n}}, \quad n \in \mathcal{N}^{\text{DL}},
	\end{split}
\end{equation}
\begin{equation} \label{eq:ul_freq_i2}
	\begin{split}
		z_{b,n} & = \eexpo{-j\theta_{a}^{\text{DL} }} y_{b,n} = \eabsn{h_{a}^{\text{UL}}}{2} x_{u,n} \\
		&  + \eexpo{-j\theta_{a}^{\text{DL}}} \ecs{\eabsn{h_{e,n}}{2} x_{e,u,n} + w_{b,n}}, \quad n \in \mathcal{N}^{\text{UL}},
	\end{split}
\end{equation}
respectively. Note that the combining corresponds to a phase rotation which does not enhance the noise and interference.

The proposed precoders, given in (\ref{eq:v_dl_i1})-(\ref{eq:v_i3}), make the eavesdropping process more difficult since Eve must search two different precoders, enlarging the execution time of the optimization algorithm. To fully avoid eavesdropping, the common phase response of the passive elements at D-RIS ($\varphi_{a}^{\text{DL}}, \varphi_{a}^{\text{UL}}$) can be changed from different slots. Therefore, even if Eve was able to find them out, the precoders would be expired and useless.

The use of additional terms $v_{b,n}^{\text{UL}}$ and $v_{u,n}^{\text{DL}}$ in the computation of precoding matrices will force the receiver to perform an additional combining, given in (\ref{eq:dl_freq_i2})-(\ref{eq:ul_freq_i2}). This will allow the receiver not only to check the identity of the transmitter but also avoid the demodulation of the injected false data stream reflected by Eve ($x_{e,b,n}$ and $x_{e,u,n}$) since the combiner is rotating these fake symbols, regardless the amount of signal strength of both original and fake data streams. Consequently, the false data injection by Eve is even more difficult than the eavesdropping process, since Eve not only has to discover the two precoders ($v_{b,n}$ and $v_{u,n}$), but also must find out the terms $v_{b,n}^{\text{UL}}$ and $v_{u,n}^{\text{DL}}$, which is additionally enhancing the difficulty of the optimization problem.

\eAddTwoColFig{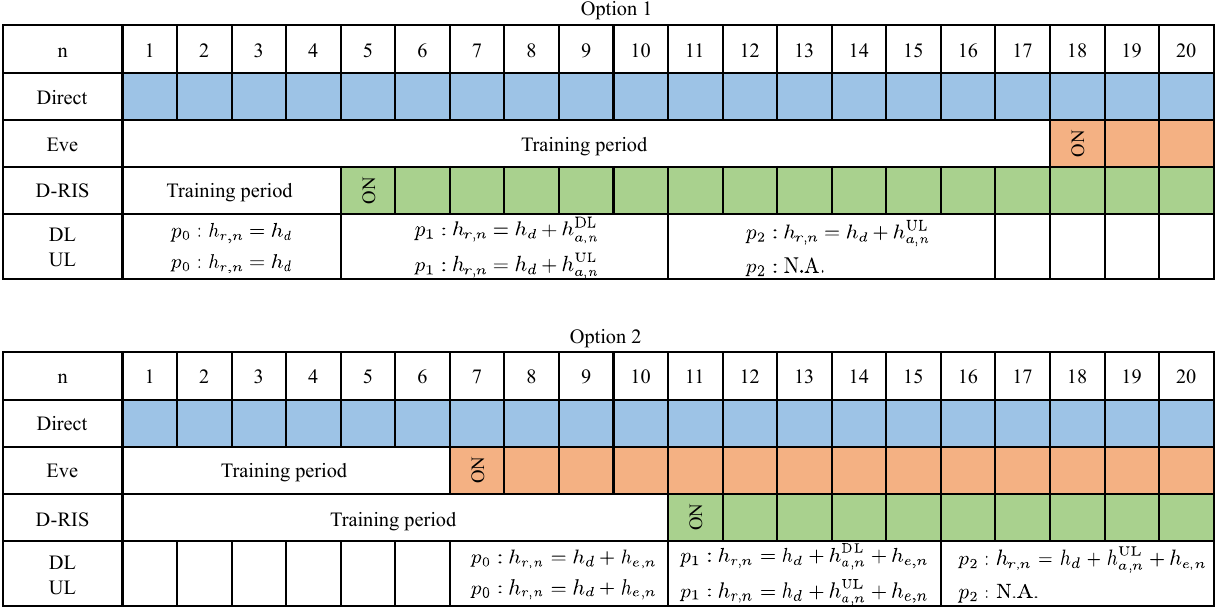}{1}{fig:chan_best}{Illustration of the CEP for an interference-free scenario. It can estimate the CSI of both DL and UL between BS and UE via D-RIS. In option 1, the D-RIS is first activated, meanwhile, in option 2, Eve is first activated. In both options, the CEP can retrieve the desired CSI of the D-RIS by linearly combining the estimated channel of the three phases ($p_{0}, p_{1}, p_{2}$).}

\subsection{Polluted CSI of D-RIS Link}
\label{sec:anti2}
In this case, the legitimate entities (BS and UE) cannot obtain the CSI of the non-reciprocal channel via D-RIS (neither $h_{a}^{\text{DL}}$ nor $h_{a}^{\text{UL}}$) by using the CEP. The main reason for this situation is that Eve is more advanced and interferes constantly with the CEP of the D-RIS, such that the obtained estimates are polluted by the CSI of Eve ($h_{e,n}$), as detailed in Section \ref{sec:chan}. By considering the polluted CSI of the D-RIS, (\ref{eq:v_dl_i1})-(\ref{eq:v_i3}) are replaced by
\begin{equation} \label{eq:v_dl_w1}
	v_{b,n} = \econj{h_{a}^{\text{DL}}+h_{e,n}} \eexpo{j\eangle{h_{a}^{\text{UL}}+h_{e,n}}}, \quad n \in \mathcal{N}^{\text{DL}},
\end{equation}
\begin{equation} \label{eq:v_ul_w1}
	v_{b,n} = \econj{h_{a}^{\text{UL}}+h_{e,n}} \eexpo{j\eangle{h_{a}^{\text{DL}}+h_{e,n}}}, \quad n \in \mathcal{N}^{\text{UL}}, 
\end{equation}
respectively. Given these precoders, none of the three entities (BS, UE and Eve) can satisfactorily demodulate the received symbols since the precoders are not able to either compensate for the effects of the non-reciprocal channels via D-RIS or verify the identity of the transmitter by the receiver. Consequently, the higher layers should launch again the CEP due to the presence of a high number of errors in the detection. From a communication perspective, this situation may be undesirable since the data transmission between legitimate entities is blocked. However, from the perspective of security, the proposed system successfully detected the presence of Eve, and therefore both data leakage of the transmitted symbols and false data injection at the received symbols are fully avoided, thanks to the proposed precoding and combining techniques, respectively.

\section{ Non-Reciprocal Channel Estimation Procedure (CEP) based on Phase Flipping at RIS }
\label{sec:chan}

In this section, the CEP for non-reciprocal channels via D-RIS is detailed. Note that even though the system is TDD, where the DL and UL transmissions are scheduled in different OFDM symbols within one slot, however, this system is behaving as a \acrfull{fdd} link \cite{mimo_fdd} from the perspective of CSI acquisition since the channel responses of the DL and UL are different as a consequence of the D-RIS, unlike the direct and Eve's channels.

The CSI in FDD is typically acquired at the receiver thanks to the transmission of some reference signals. In the particular case of DL, the UE must feed this CSI back to the BS and let it perform both the combining in UL and precoding in DL. Consequently, the UE does not have to perform any additional operation, trusting that the legitimate BS has correctly performed the precoding. However, in the presence of Eve, she will be able to intercept the CSI sent by the UE to obtain the used precoding matrix at the transmitter. In the case that the CSI feedback was encrypted, Eve would break it. Moreover, stronger encryption methods would require more resources in terms of time and energy needed to perform the encryption/decryption process. Then, Eve will easily eavesdrop and inject false data symbols into the original data stream and trick the receiver as explained in Section \ref{sec:attack}, seriously compromising the system. The proposed CEP based on phase flipping at the RIS is capable of avoiding the vulnerability of the CSI feedback in the CSI acquisition of the DL and UL channels for both BS and UE. The common dynamic phase, given in (\ref{eq:tvphase2}), can be exchanged between the DL and UL OFDM symbols to allow the estimation of the channel frequency response of the UL by the UE. Meanwhile, the BS can internally compute the CSI of the DL since it handles the RIS, according to Section \ref{sec:sys}, and it has both values of the common dynamic phases ($\varphi_{a}^{\text{DL}}$ and $\varphi_{a}^{\text{UL}}$). In the case that the UE was the entity that manages the D-RIS, the described CEP could be straightforwardly changed. With the availability of the CSI, both BS and UE will be able to perform both the precoding and combining techniques given in Section \ref{sec:anti}.

\subsection{Phase Flipping at D-RIS}

Given the resource sets for DL and UL ($\mathcal{N}^{\text{DL}}$ and $\mathcal{N}^{\text{UL}}$, respectively), they can be split into the following subsets as
\begin{equation} \label{eq:setn3}
	\mathcal{N}^{\text{DL}} = \mathcal{N}_{d}^{\text{DL}} \cup \mathcal{N}_{p_0}^{\text{DL}} \cup \mathcal{N}_{p_1}^{\text{DL}} \cup \mathcal{N}_{p_2}^{\text{DL}},
\end{equation}
\begin{equation} \label{eq:setn4}
	\mathcal{N}_{d}^{\text{DL}} \cap \mathcal{N}_{p_0}^{\text{DL}} \cap \mathcal{N}_{p_1}^{\text{DL}} \cap \mathcal{N}_{p_2}^{\text{DL}}  = \emptyset,
\end{equation}
\begin{equation} \label{eq:setn5}
\mathcal{N}^{\text{UL}} = \mathcal{N}_{d}^{\text{UL}} \cup \mathcal{N}_{p_0}^{\text{UL}} \cup \mathcal{N}_{p_1}^{\text{UL}}, \quad \mathcal{N}_{d}^{\text{UL}} \cap \mathcal{N}_{p_0}^{\text{UL}} \cap \mathcal{N}_{p_1}^{\text{UL}} = \emptyset,
\end{equation}
where $\mathcal{N}_{d}^{\text{DL}} $ and $\mathcal{N}_{d}^{\text{UL}} $ are the subsets that contain the OFDM symbol indices for the transmission of the data symbols in DL and UL, respectively. $\mathcal{N}_{p_i}^{\text{DL}}$, $0\leq i \leq 2$ and $\mathcal{N}_{p_i}^{\text{UL}}$, $0\leq i \leq 1$, correspond to the subset of OFDM symbol indices for the transmission of pilots symbols at $i$-th stage in the DL and UL, respectively. 

In the first stage ($p_{0}$), the D-RIS is switched off allowing the obtention of the CSI of the equivalent channel between the legitimate entities without the effect of the D-RIS. Then, in the second and third stages ($p_{1}$ and $p_{2}$), the D-RIS is switched on and the common dynamic phase configuration is set as
\begin{equation} \label{eq:tvphase3}
	\varphi_{a,n} = \begin{cases} 
		\varphi_{a}^{\text{DL}} & n \in \mathcal{N}_{d}^{\text{DL}} \cup \mathcal{N}_{p_1}^{\text{DL}} \\ 
		\varphi_{a} ^{\text{UL}}  & n \in \mathcal{N}_{d}^{\text{UL}} \cup \mathcal{N}_{p_1}^{\text{UL}} \cup \mathcal{N}_{p_2}^{\text{DL}}
	\end{cases}.
\end{equation}
Firstly, the common dynamic phase configuration set for data transmission and $p_{1}$ are the same ones, and hence, the BS and UE can obtain the CSI of the UL and DL channels ($h_{a}^{\text{UL}}$ and $h_{a}^{\text{DL}}$), respectively. Then, the chosen phase configurations between DL and UL are exchanged in $p_{2}$ for the particular case of DL. This phase flipping at the D-RIS will allow the UE to easily estimate the CSI of the UL ($h_{a}^{\text{UL}}$), meanwhile the BS can straightforwardly obtain the CSI of the DL by performing
\begin{equation} \label{eq:bs_compu}
	h_{a}^{\text{DL}}=h_{a}^{\text{UL}}\eexpo{j\ecs{\varphi_{a}^{\text{DL}}-\varphi_{a} ^{\text{UL}}}},
\end{equation}
because the channel between BS$\leftrightarrow$RIS and RIS$\leftrightarrow$UE are reciprocal, as stated in (\ref{eq:ha})-(\ref{eq:ha_distri}). Finally, the BS and UE have the CSI of both DL and UL, allowing the execution of the precoding and combining processes described in Section \ref{sec:anti}. Moreover, note that the requirement of feeding back of the estimated CSI is fully avoided thanks to the phase flipping procedure.

\eAddTwoColFig{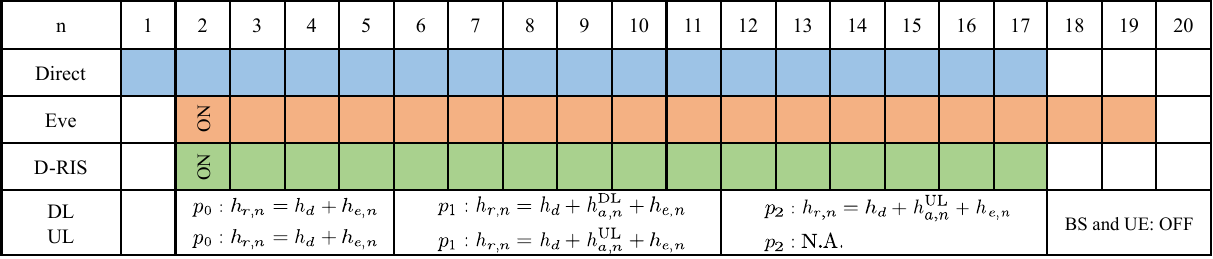}{1}{fig:chan_worst}{Illustration of the scenario B. The CEP is not able to successfully obtain the CSI of the D-RIS channel since both D-RIS and Eve are activated at the same moment. }

It is considered that both the adversarial RIS equipped by Eve and D-RIS are activated once they have found out their respective best phase configurations and they are capable of providing an alternative path between the BS$\leftrightarrow$UE. Otherwise, it is assumed that either Eve or D-RIS is deactivated, which means that the panel is either switched off or they are in their training period seeking their phase configurations. Additionally, according to Section \ref{sec:anti}, two possible scenarios are considered due to the presence of Eve, which are detailed in the following subsections.

\subsection{Interference-free CSI of the D-RIS Link}

Fig. \ref{fig:chan_best} plots two possible options for this particular scenario. In the beginning, both Eve and D-RIS are looking for their best phase configurations (training period), while the original link is the only way to exchange the information between the legitimate entities, and hence the CSI of the direct channel can be satisfactorily obtained. Then, the D-RIS and Eve are activated in different moments of the slot. In option one, the CSI of the D-RIS can be obtained in the second stage since D-RIS is activated before Eve. In option two, the CSI of D-RIS is retrieved in the third stage since the D-RIS needs more time to find out its best phase configurations. The CEP can successfully retrieve the CSI of all channels by solving some straightforward linear equations.

Focusing on the CEP in the DL, the estimated CSI obtained at the UE in option one are
\begin{equation} \label{eq:est_chan_o1p0}
	\widehat{h}_{r,n} = h_{d} + w'_{u,n}, \quad n \in \mathcal{N}_{p0}^{\text{DL}},
\end{equation}
\begin{equation} \label{eq:est_chan_o1p1}
	\widehat{h}_{r,n} = h_{d} + h_{a,n}^{\text{DL}} + w'_{u,n}, \quad n \in \mathcal{N}_{p1}^{\text{DL}},
\end{equation}
\begin{equation} \label{eq:est_chan_o1p2}
	\widehat{h}_{r,n} = h_{d} + h_{a,n}^{\text{UL}} + w'_{u,n}, \quad n \in \mathcal{N}_{p2}^{\text{DL}},
\end{equation}
where $w'_{u,n}$ is the noise term after applying the LS at the UE in the $n$-th OFDM symbol and a particular subcarrier. The power of the pilot symbol is normalized to one as data symbols, the distribution of $w'_{u,n}$ is also $\egausd{0}{\sigma_{w}^{2}}$. By combining (\ref{eq:est_chan_o1p0})-(\ref{eq:est_chan_o1p2}), the UE can obtain the CSI of D-RIS channels for both DL and UL. Alternatively, the estimated CSI obtained at the UE in option two are
\begin{equation} \label{eq:est_chan_o2p0}
	\widehat{h}_{r,n} = h_{d} + h_{e,n}+ w'_{u,n}, \quad n \in \mathcal{N}_{p0}^{\text{DL}},
\end{equation}
\begin{equation} \label{eq:est_chan_o2p1}
	\widehat{h}_{r,n}= h_{d} + h_{a,n}^{\text{DL}} + h_{e,n} + w'_{u,n}, \quad n \in \mathcal{N}_{p1}^{\text{DL}},
\end{equation}
\begin{equation} \label{eq:est_chan_o2p2}
	\widehat{h}_{r,n} = h_{d} + h_{a,n}^{\text{UL}} + h_{e,n} + w'_{u,n}, \quad n \in \mathcal{N}_{p2}^{\text{DL}}.
\end{equation}
Again, the CSI of the DL and UL provided by the D-RIS can be obtained by linearly combining (\ref{eq:est_chan_o2p0})-(\ref{eq:est_chan_o2p2}) at the UE. 

Note that the CEP described in the DL is the same procedure for the UL. The CSI acquisition at the BS can be analogously obtained by using (\ref{eq:est_chan_o1p0})-(\ref{eq:est_chan_o1p1}) and (\ref{eq:est_chan_o2p0})-(\ref{eq:est_chan_o2p1}) to obtain $\widehat{h}_{a,n}^{\text{UL}}$, while $\widehat{h}_{a,n}^{\text{DL}}$ can be computed by using (\ref{eq:bs_compu}).

\subsection{Polluted CSI of D-RIS Link}

\eAddTwoColFig{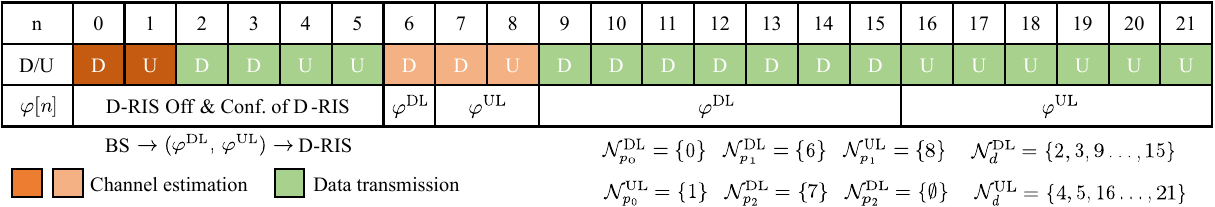}{1}{fig:chan_slots}{Example of OFDM symbol allocation for performing non-reciprocal channel estimation and data transmission.}

Fig. \ref{fig:chan_worst} illustrates the undesirable situation in which a smart Eve constantly interferes with the CEP by activating herself at the same moment as the D-RIS. To avoid this issue, the D-RIS should be deactivated to allow the estimation of the CSI of Eve. Then, after solving a linear equation, both CSI from the direct and D-RIS channels are available.

On one hand, focusing on the DL, the CSI obtained at the UE is given by
\begin{equation} \label{eq:est_chan_o3p0}
	\widehat{h}_{r,n} = h_{d} + h_{e} + w'_{u,n}, \quad n \in \mathcal{N}_{p0}^{\text{DL}},
\end{equation}
\begin{equation} \label{eq:est_chan_o3p1}
	\widehat{h}_{r,n}= h_{d} + h_{a,n}^{\text{DL}} + h_{e,n} + w'_{u,n}, \quad n \in \mathcal{N}_{p1}^{\text{DL}},
\end{equation}
\begin{equation} \label{eq:est_chan_o3p2}
	\widehat{h}_{r,n} = h_{d} + h_{a,n}^{\text{UL}} + h_{e,n} + w'_{u,n}, \quad n \in \mathcal{N}_{p2}^{\text{DL}}.
\end{equation}
On the other hand, focusing on the DL, the CSI obtained at the BS is given by
\begin{equation} \label{eq:est_chan_o4p0}
	\widehat{h}_{r,n} = h_{d} + h_{e} + w'_{u,n}, \quad n \in \mathcal{N}_{p0}^{\text{UL}},
\end{equation}
\begin{equation} \label{eq:est_chan_o4p1}
	\widehat{h}_{r,n}= h_{d} + h_{a,n}^{\text{UL}} + h_{e,n} + w'_{u,n}, \quad n \in \mathcal{N}_{p1}^{\text{UL}},
\end{equation}

After combining (\ref{eq:est_chan_o3p0})-(\ref{eq:est_chan_o3p2}), (\ref{eq:est_chan_o4p0})-(\ref{eq:est_chan_o4p1}) and (\ref{eq:bs_compu}), the CSI of the D-RIS channel at the UE and BS, respectively, cannot be successfully retrieved due to the interference produced by Eve. In this case, the described precoding and combining procedures given in Subsection \ref{sec:anti2} should be able to detect this situation. Then, the transmitter should be muted for some random number of OFDM symbols until Eve is switched off. Finally, the CEP should be restarted again.

\subsection{Proposed Slot Structure and Implementation Aspects}

Following the 5G standard \cite{nr-211} and considering the requirements of the proposed CEP, the \acrfull{dmrs} can be exploited for this purpose with some modifications. Since up to four OFDM symbols can be activated within one slot of seven OFDM symbols, the proposed CEP can be implemented by combining the CSI obtained from, at least, two slots. Hence, the channel coherence time should last, at least, these two slots. However, since all the entities are fixed in their respective static place, the coherence time may cope with more slots. Additionally, the resource allocation of DM-RS within one slot should be changed from the current slot to the following one. It is crucial to keep in secret the indices of those OFDM symbols in the slot that are carrying the DM-RS for performing the CEP and prevent potential disruption to the CEP from any illegitimate intruder. Hence, the resource allocation of the control channels should be pseudo-randomly changed for each slot.

An example of the resource allocation of one proposed slot for transmitting both reference signals and data symbols in both DL and UL is plotted in Fig. \ref{fig:chan_slots}. To deploy the CEP, the total number of OFDM symbols required for transmitting reference signals for DL and UL in one slot, given in (\ref{eq:tvphase3}), corresponds to $\eabsn{\mathcal{N}_{p}^{\text{DL}}}{}=2$ and $\eabsn{\mathcal{N}_{p}^{\text{UL}}}{}=3$ out of $\eabsn{\mathcal{N}}{}=22$. Note that the proposed CEP for a non-reciprocal channel only requires one additional OFDM symbol for transmitting the reference signals as compared to the reciprocal one, to execute the phase flipping technique, which corresponds to a negligible impact in the data rate.

Assuming that the BS is responsible for configuring and managing the D-RIS, the former transmits to the latter the common time-varying phases per slot ($\varphi_{a}^{\text{DL}}$ and $\varphi_{a}^{\text{UL}}$) in advance, for example during the training phase of the D-RIS. Note that, the duration of the transmission of these two phases is negligible as compared to either the time required for the training or the duration of the slot. Later, these two phase values can be programmed to be automatically configured once the D-RIS is on in those scheduled OFDM symbols within one slot (see Fig. \ref{fig:chan_slots}). It is assumed that the configuration time of these two common phases is shorter than the duration of the CP. Otherwise, an additional guard interval is required at the expense of a slight reduction in the efficiency of the system. Finally, to keep the security of the control signal transmission against interception and false data injection, all the signalling between the BS and D-RIS should be encoded. In the case that the D-RIS did not have enough computational resources for encoding all the control messages, at least, these two common time-varying phases must be encoded since they represent the key information used to protect the data transmission.

\section{Analysis of the Achievable Secrecy Rate (ASR) and Probability of Fake Symbol Detection}
\label{sec:sar}

This section provides the analysis of the performance of the proposed defensive system under RITM attack based on non-reciprocal channels via D-RIS. On one hand, the ASR is analysed for the eavesdropping stage, it accounts for the amount of information eavesdropped by Eve as compared to the information received by the legitimate receiver. On the other hand, the probability of fake data detection is obtained for the false data injection stage, it measures the probability of Eve to trick the legitimate receiver to demodulate the fake data. Additionally, its performance is also compared to the hypothetical case based on the traditional reciprocal channel via RIS, showing that our proposed system is significantly better.

\subsection{Eavesdropping: ASR}
In this case, Eve is only listening to the data information transmitted by both legitimate entities of the link (BS and UE). The achievable rate of any communication of interest can be obtained as
\begin{equation} \label{eq:achi01}
	C_{i} = \eta_{i}\elogtwo{1+\rho_{i}}, \quad i \in \ebs{d,e,a}, \quad \eta_{i} = 1 - \frac{N_{p}}{N},
\end{equation}
where $N_{p}$ is the number of OFDM symbols devoted to the transmission of reference signals, $\eta_{i}$ and $\rho_{i}$ correspond to the efficiency and SNR of the link of interest, respectively, and $i$ is a token.

Then, the ASR \cite{sec} between the direct and Eve links measures the amount of information eavesdropped by Eve as compared to the legitimate link, which can be defined as
\begin{equation} \label{eq:eve01}
	E_{d} = \begin{cases} 
		C_{d} - C_{e}  & \rho_{d} > \rho_{e} \\ 
		0 & \rho_{d} \leq \rho_{e}
	\end{cases}, \quad \eta_{d} = \eta_{e}, \quad \rho_{e} \in \ebs{\rho_{eb},\rho_{eu}},
\end{equation}
\begin{equation} \label{eq:snr_d}
	\rho_{d} = \frac{\eexpabstwo{h_{d}}}{\sigma_{w}^{2}} = \frac{\sigma_{d}^{2}}{\sigma_{w}^{2}}, \quad \eexpabsn{x_{b,n}}{2} = \eexpabsn{x_{u,n}}{2} = 1,
\end{equation}
\begin{equation} \label{eq:snr_heb}
	\rho_{eb} = \frac{\eexpabstwo{h_{e,b,n}}}{\sigma_{w}^{2}} = M_{e} \frac{\sigma_{q_{e}}^{2}\sigma_{g_{v}}^{2}}{\sigma_{w}^{2}},
\end{equation}
\begin{equation} \label{eq:snr_heu}
	\rho_{eu} = \frac{\eexpabstwo{h_{e,u,n}}}{\sigma_{w}^{2}} = M_{e} \frac{ \sigma_{g_{e}}^{2} \sigma_{g_{v}}^{2} }{\sigma_{w}^{2}},
\end{equation}
where $\rho_{d}$ is the SNR of the direct link between BS$\leftrightarrow$UE, while $\rho_{eq}$ and $\rho_{eg}$ correspond to the SNRs of the links between BS$\leftrightarrow$Eve and UE$\leftrightarrow$Eve, respectively. Regarding the efficiency of the system, $N_{p} = 1$ since the CSI can be obtained in the UL and reused in the DL due to the channel reciprocity property. Note that the average gain of the cascaded channel of the adversarial RIS equipped by Eve is obtained by approximating its probability density function as a normal distribution, making use of the \acrfull{clt} \cite{clt}. This approximation can be also employed for the D-RIS.

Typically, $E_{d}=0$, because the SNR of Eve is always better than the direct link since the channel propagation of the former is better than the latter. Moreover, the high number of passive elements of the RIS, equipped by Eve, will enhance the received signal of the link. To circumvent this issue, the D-RIS is introduced to provide also an alternative link between the legitimate entities with a higher SNR, and hence, the ASR and its corresponding SNR can be obtained as
\begin{equation} \label{eq:eve02}
	E_{a} = \begin{cases} 
		C_{a} - C_{e}  & \rho_{a} > \rho_{e} \\ 
		0 & \rho_{a} \leq \rho_{e}
	\end{cases}, \quad \eta_{a} = \eta_{e}, \quad \rho_{e} \in \ebs{\rho_{eb},\rho_{eu}},
\end{equation}
\begin{equation} \label{eq:snr_a}
	\rho_{a} = \frac{\eexpabstwo{h_{a,n}}}{\sigma_{w}^{2}} = M_{a} \frac{\sigma_{q_{a}}^{2}\sigma_{g_{a}}^{2} }{\sigma_{w}^{2}},
\end{equation}
respectively. By comparing (\ref{eq:eve01})-(\ref{eq:snr_heu}) with (\ref{eq:eve02})-(\ref{eq:snr_a}), it points out that increasing the number of passive elements of D-RIS is not the only way to enhance the $C_{a}$. As an alternative, minimizing $C_{e}$ will also help mitigate the information eavesdropped by Eve.

To avoid Eve accessing the information transmitted by the legitimate links, precoding is required to mask the transmitted information. In the case of using a reciprocal channel provided D-RIS, $N_{p}=4$ since both BS and UE must transmit reference signals to allow the estimation of the CSI without the need of feeding back it. Consequently, (\ref{eq:eve02}) can be rewritten as
\begin{equation} \label{eq:eve03}
	E_{ar} = \frac{N_{r}}{N}C_{a} + \ecs{1-\frac{N_{r}}{N}} \ecs{C_{a}-C_{e}} = C_{a} - \ecs{1-\frac{N_{r}}{N}} C_{e},
\end{equation}
where $E_{ar}$ is the ASR of a reciprocal channel, $N_{r}$ is the amount of time measured in OFDM symbols that Eve requires to find out the used precoder or the CSI of the reciprocal channel by the legitimate links. On one hand, the first term of (\ref{eq:eve03}) accounts for the ASR of the link when Eve has not discovered the used precoders yet, and hence, she is not able to obtain any information. On the other hand, the second term of (\ref{eq:eve03}) points out the ASR once Eve has found out the precoders, and therefore, the term $C_{e}$ is penalizing the expression. 

According to our proposed non-reciprocal channel ($N_{p}=3$), (\ref{eq:eve03}) can be further improved by
\begin{equation} \label{eq:eve04}
	E_{an} = C_{a} - \ecs{1-\frac{N_{n}}{N}} C_{e},
\end{equation}
where $E_{an}$ is the ASR of a non-reciprocal channel, $N_{n}$ is the amount of time measured in OFDM symbols that Eve requires to find out the used precoders by the legitimate links in a non-reciprocal channel. According to (\ref{eq:v_dl_i1})-(\ref{eq:v_i3}), Eve must search two different precoders to allow her to eavesdrop on the transmitted information by both BS and UE. Hence, the proposed system based on a non-reciprocal channel, given in (\ref{eq:eve04}), doubles the required searching time of Eve to the reciprocal case given in (\ref{eq:eve03}) ($N_{n} = 2 N_{r}$).

The final expression for the ASR for the non-reciprocal channel for the D-RIS can be expressed as
\begin{equation} \label{eq:eve05}
	E_{an} = \eta_{a}\ecs{\elogtwo{1+\rho_{a}} - \ecs{1-\frac{N_{n}}{N}} \elogtwo{1+\rho_{e}}},
\end{equation}
where $\rho_{e}\in\ebs{\rho_{eb},\rho_{eu}}$. Since the SNR of the D-RIS and Eve's link is higher than the unity since the number of passive elements of the D-RIS and Eve is extremely large, (\ref{eq:eve05}) can be approximated by
\begin{equation} \label{eq:eve06}
	E_{an} \approx \eta_{a}\ecs{\elogtwo{\rho_{a}} - \ecs{1-\frac{N_{n}}{N}} \elogtwo{\rho_{e}}}.
\end{equation}
Substituting (\ref{eq:snr_heb}), (\ref{eq:snr_heu}) and (\ref{eq:snr_a}) in (\ref{eq:eve06}), the ASR between the D-RIS link and the link between BS$\leftrightarrow$Eve and UE$\leftrightarrow$Eve is approximated by
\begin{equation} \label{eq:eve07}
	E_{an}  \approx \eta_{a}  \elogtwo{\frac{M_{a}^{2}\sigma_{q_{a}}^{2}\sigma_{g_{a}}^{2}}{\ecs{M_{e}\sigma_{e}^{2}\sigma_{g_{v}}^{2}}^{1-\frac{N_{n}}{N}}\ecs{\sigma_{w}^{2}}^{\frac{N_{n}}{N}}}},
\end{equation}
\begin{equation*}
	\sigma_{e}^{2} \in \ebs{\sigma_{q_{e}}^{2},\sigma_{g_{e}}^{2}},
\end{equation*}
respectively. To minimize the eavesdropped information by Eve, it should satisfy that
\begin{equation} \label{eq:effs}
	1-\frac{N_{n}}{N} \leq 0 \rightarrow N \leq N_{n},
\end{equation}
where it points out that the number of OFDM symbols within one slot should be, at maximum, the required time for obtaining the two precoders. Additionally, taking into account the efficiency of the system given in (\ref{eq:achi01}), it should satisfy that
\begin{equation} \label{eq:efft}
	N_{p} < N \leq N_{n},
\end{equation}
where the number of OFDM symbols within one slot should be larger than the number of OFDM symbols devoted for the transmission of the reference symbols, and at the same time, lower than the execution time for finding both DL and UL precoders by Eve. In realistic scenarios, (\ref{eq:efft}) can be easily satisfied since not only optimization problems require a significant amount of time, but also their networking and processing times for transferring the received signals from the physical layer to the higher layers.

\subsection{Data Manipulation: Probability of Fake Symbol Detection}

Later, Eve can also inject false data into all legitimate receivers. Hence, this receiver is receiving both the original and false symbols sent by the legitimate transmitter and Eve, respectively. Depending on the signal strength of both signals and the noise, the receiver may wrongly decode the fake signal injected by Eve instead of the original one.

The probability of fake symbol detection in the proposed system can be defined as
\begin{equation} \label{eq:prob_res}
	P_{r} = \frac{N'_{n}}{N} P_{1} + \ecs{1-\frac{N'_{n}}{N}} P_{2} ,
\end{equation}
where $P_{1}$ and $P_{2}$ are the probabilities of decoding the fake message sent by Eve before and after finding out the employed combiners, respectively. $N'_{n}$ accounts for the time required by Eve not only to obtain the used precoders by BS ($v_{b,n}$) and UE ($v_{u,n}$), but she must also compute the two terms of each precoder to find out the combiners ($v_{b,n}^{\text{UL}}$ and $v_{u,n}^{\text{DL}}$). Therefore, the proposed defensive system is even more robust against false data injection rather than eavesdropping since it satisfies that $N'_{n}>>N_{n}$.

It is clear that before finding out the combiner, the legitimate receiver will not be able to decode the false message ($P_{1}=0$). Hence, (\ref{eq:prob_res}) can be simplified as
\begin{equation} \label{eq:prob_res2}
	P_{r} = \ecs{1-\frac{N'_{n}}{N}} P_{2}.
\end{equation}
To demodulate the injected signal by Eve at any legitimate receiver, its power should be higher than the power threshold, and hence the resilience probability can be defined as
\begin{equation} \label{eq:prob_res3}
		P_{2} = \eprob{\eabsn{h_{e,n}}{2} > \beta},
\end{equation}
where $\beta$ is the power threshold, and given the scenario depicted in Fig. \ref{fig:system}, it can be calculated as 
\begin{equation} \label{eq:beta}
	\beta = M_{a}\sigma_{q_{a}}^{2}\sigma_{g_{a}}^{2} + \sigma_{d}^{2}+\sigma_{w}^{2} ,
\end{equation}
where the first term accounts for the average gain of the cascaded channel of D-RIS, obtained in (\ref{eq:snr_a}). Therefore, substituting (\ref{eq:beta}) in (\ref{eq:prob_res}), the resilience probability can be computed as
\begin{equation} \label{eq:prob_res4}
	\begin{split}
		P_{2} & = \eprob{\eabsn{h_{e,n}}{2} > M_{a}\sigma_{q_{a}}^{2}\sigma_{g_{a}}^{2}+\sigma_{d}^{2}+\sigma_{w}^{2}} \\
		& = \eexpo{-\frac{M_{a}\sigma_{q_{a}}^{2}\sigma_{g_{a}}^{2}+\sigma_{d}^{2}+\sigma_{w}^{2}}{M_{e}\sigma_{e}^{2}\sigma_{g_{v}}^{2}}},
	\end{split}
\end{equation}
\begin{equation*}
	\sigma_{e}^{2} \in \ebs{\sigma_{q_{e}}^{2},\sigma_{g_{e}}^{2}}.
\end{equation*}

Finally, substituting (\ref{eq:prob_res4}) in (\ref{eq:prob_res2}), the probability of detecting fake symbols can be rewritten as
\begin{equation} \label{eq:prob_res5}
	P_{r} =\ecs{1-\frac{N'_{n}}{N}} \eexpo{-\frac{M_{a}\sigma_{q_{a}}^{2}\sigma_{g_{a}}^{2}+\sigma_{d}^{2}+\sigma_{w}^{2}}{M_{e}\sigma_{e}^{2}\sigma_{g_{v}}^{2}}},
\end{equation}
\begin{equation*}
	\sigma_{e}^{2} \in \ebs{\sigma_{q_{e}}^{2},\sigma_{g_{e}}^{2}}.
\end{equation*}
Note that for the reciprocal case, since there is not any additional combining procedure capable of checking the integrity of the received message, its resilience probability is $P_{r} = 1-P_{2}$, which has significantly worse performance than the proposed technique.

\section{Performance Evaluation}
\label{sec:per}

In this section, several numerical results are provided to show the performance of the proposed defensive system against RITM attack, based on the exploitation of non-reciprocal channels via D-RIS. In this performance comparison, the traditional reciprocal channel via regular RIS is considered as the baseline. A summary of the simulation parameters is given in Table \ref{tab:simparam}. The employed slot structure corresponds to Fig. \ref{fig:chan_slots}. The channel propagation model adopted corresponds to the factory scenario given in \cite{nr-901}. The direct channel between BS$\leftrightarrow$UE, the channels via the D-RIS (BS$\leftrightarrow$D-RIS and UE$\leftrightarrow$D-RIS) and adversarial RIS (BS$\leftrightarrow$adversarial RIS, UE$\leftrightarrow$adversarial RIS and Eve$\leftrightarrow$adversarial RIS) are assumed to be NLOS since some obstacles may obstruct the ray. The geographical location of all entities is specified in Cartesian coordinates $\ecs{x,y,z}$ m. Additionally, let us define $\eta_{s}$ which is the percentage of the slot that Eve is jeopardizing the legitimate link since she has already found out the precoders/combiners, and it can be defined as
\begin{equation} \label{eq:eff}
	\eta_{s} = \begin{cases} 
		1 - N_{r} / N & \text{Reciprocal chan.}\\ 
		1 - N_{n} / N  & \text{Non-recipr. chan. (eavesdropping)} \\
		1 - N'_{n} / N  & \text{Non-recipr. chan. (data manipulation)}
	\end{cases}.
\end{equation}

Note that the existing solutions for preventing Eve given in \cite{riseve3,riseve4,riseve5} cannot be fairly compared to the proposal since they are not only assuming that the legitimate entities are always aware of the existence of Eve, but they also know some other relevant information about this adversary, such as her (instantaneous) CSI and geographical position. These considerations are unrealistic because this information related to Eve will never be available, as shown in \cite{ristest1,ristest2}. Later, given the availability of the CSI and the geographical positions of both legitimate entities and Eve, these methods also require a large amount of time to solve complex optimization problems to find a pair of precoders and combiners capable of avoiding Eve in a multi-antenna system. On the contrary, the proposed defensive technique considers that no information about Eve is available, and a single antenna is considered to be equipped by both legitimate entities. Additionally, the precoders and combiners exploited by the proposed solution can be efficiently obtained by combining the CSI of DL and UL, significantly reducing the computation time.

\begin{table}[!t]
	\centering
	\caption{Simulation Parameters}
	\label{tab:simparam}
	\begin{tabular}{|c|c|c|c|c|c|}
		\hline
		\textbf{BS loc.}    & $\ecs{0,0}$     & $\mathbf{P_ {\textbf{max}}}$    & $-30$ dBm       & $\mathbf{M_{e}}$          & $1000$       \\ \hline
		\textbf{UE loc.}    & $\ecs{20,0}$    & $\mathbf{\eta_{a}}$ 		      & $0.82, 0.86$      & $\mathbf{\eta_{d}}$ 		& $0.91$       \\ \hline
		\textbf{D-RIS loc.} & $\ecs{10,5}$    & $\mathbf{\eta_{s}}$             & $0.05 - 0.6$  & $\mathbf{K}$              & $600$            \\ \hline
		\textbf{Adv. RIS.}   & $\ecs{10,-5}$   & \textbf{Eve loc.}   & $\ecs{10,-10}$  & $\mathbf{N}$              & $22$          \\ \hline
	\end{tabular}
\end{table}

\subsection{Evaluation of Eavesdropping}

In Fig. \ref{fig:resEveAchi}, the achievable rate of the different links involved in the proposed scenario in the system model is shown. These results correspond to the best performance of each link, which corresponds to an upper bound. Note that due to geometrical symmetry of the BS and UE with respect to the D-RIS and adversarial RIS (see Table \ref{tab:simparam}), the path-loss of the channels $q_{a,m}$, $g_{a,m}$, $q_{e,m}$ and $g_{e,m}$ are the same value. However, the distance between Eve and her adversarial RIS is shorter than the distance between the legitimate entities and the D-RIS and adversarial RIS, then it satisfies that $g_{v,m} > q_{a,m}, g_{a,m}, q_{e,m}, g_{e,m}$. On one hand, the exploitation of an additional D-RIS is helping to enhance the performance of the link. However, this performance is reduced by considering the efficiency factor which accounts for the number of OFDM symbols devoted to reference signals. Albeit the theoretical performance of the proposed non-reciprocal case is slightly lower than the reciprocal one, it will improve the security aspect as seen later in Fig. \ref{fig:resEveSec}. On the other hand, the achievable rate of Eve is the highest one since her channel propagation conditions are better than the direct NLOS channel (BS$\leftrightarrow$UE) and the cascaded channel (BS$\leftrightarrow$D-RIS$\leftrightarrow$UE). Hence, in the case that precoding is not used (UL case with a reciprocal channel), the secrecy achievable rate will be reduced to zero between the BS and UE since Eve can fully eavesdrop on all the transmitted information by legitimate entities. 

\eAddFig{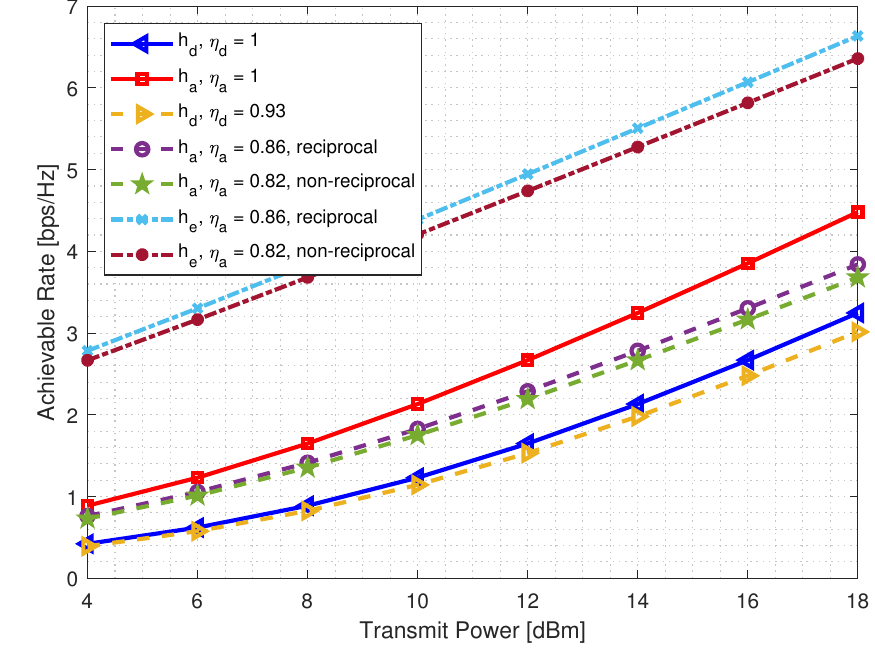}{1}{fig:resEveAchi}{Achievable rates of all the links considered a scenario in the system model for $M_{a}=2000$.}
\eAddFig{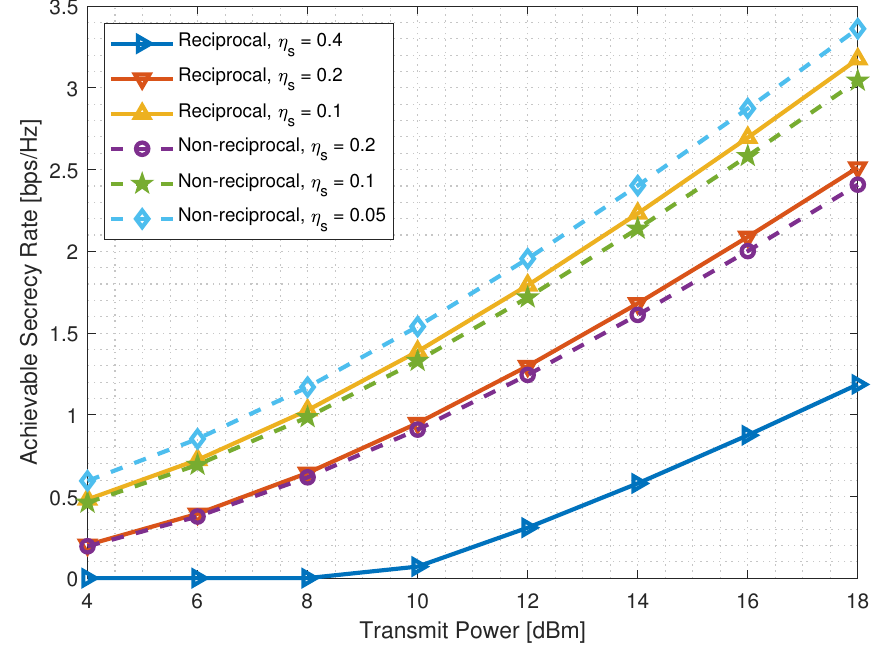}{1}{fig:resEveSec}{Comparison of the ASR for eavesdropping between reciprocal and non-reciprocal cases for $M_{a}=2000$.}

In Fig. \ref{fig:resEveSec}, the comparison of the ASR between the non-reciprocal and reciprocal channels is given by considering different values of $\eta_{s}$. It is assumed that the time required by Eve to find out the precoding matrices for the non-reciprocal case is half of the reciprocal one ($N = 2 N'$). The proposed non-reciprocal case (dotted lines) has a better performance than the reciprocal one (solid lines) when Eve is disrupting the link. Despite the proposed non-reciprocal defensive system has a lower performance than the reciprocal one from the theoretical perspective in Fig. \ref{fig:resEveAchi} due to the CEP ($\eta_{a}$), the former is more robust against attacks since the system has better protection as a consequence of using different precoders and combiners, and therefore, Eve requires more time find out the employed precoders and combiners to break the communication link.

\subsection{Evaluation of False Data Injection}

In Figs. \ref{fig:resManEff} and \ref{fig:resManSiz}, a comparison in terms of the probability of fake symbol detection between the reciprocal and non-reciprocal cases for the data manipulation scenario is given. Similarly to the previous case, it is assumed that $N = 2 N'$. In Fig. \ref{fig:resManEff}, the proposed defensive system based on a non-reciprocal channel is again always better than the reciprocal one via RIS even though it requires only one more OFDM symbol to transmit the reference signals. The reason behind these results is due to the robust design of the precoders and combiners, making the intrusion more difficult for the adversarial RIS. Note that either increasing the number of passive elements at D-RIS ($M_{a}$) or enlarging the time required to find out the precoders/combiners will lower the probability of detecting the fake symbols reflected by Eve, however, the latter is more effective than the former since it can reduce the probability by one order of magnitude. Then, in Fig. \ref{fig:resManSiz}, it verifies that when the number of passive elements at D-RIS is significantly larger than Eve ($M_{a} = 4000, 8000$ and $M_{e}=2000$), the probability of detecting the fake symbols are reduced further for both reciprocal and non-reciprocal cases as compared to smaller sizes of D-RIS. However, it is impossible to know in advance if the D-RIS has a larger number of passive elements as compared to Eve, and therefore, the proposed non-reciprocal is the safest technique which will always guarantee that the received symbols are never manipulated by Eve. Additionally, D-RIS cannot be extremely large since it requires a long training sequence \cite{ris1,ris2,ris3,ris4,ris5,ris6,ris7,riskey}, and meanwhile, Eve equipped with small size RIS can be jeopardizing the direct link.

\eAddFig{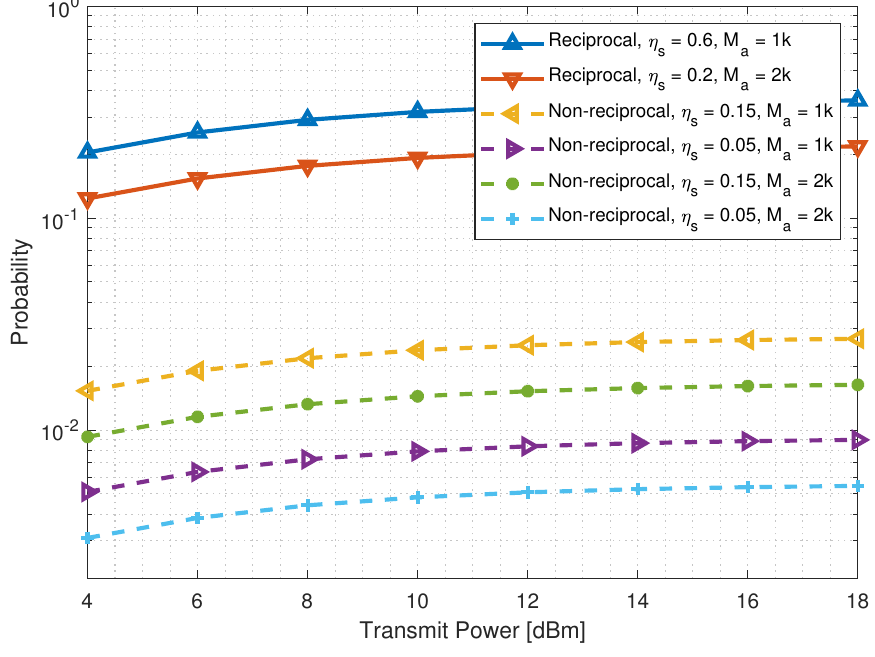}{1}{fig:resManEff}{Comparison of the probability of false data detection for different values of $\eta_{s}$ and $M_{a}$.}
\eAddFig{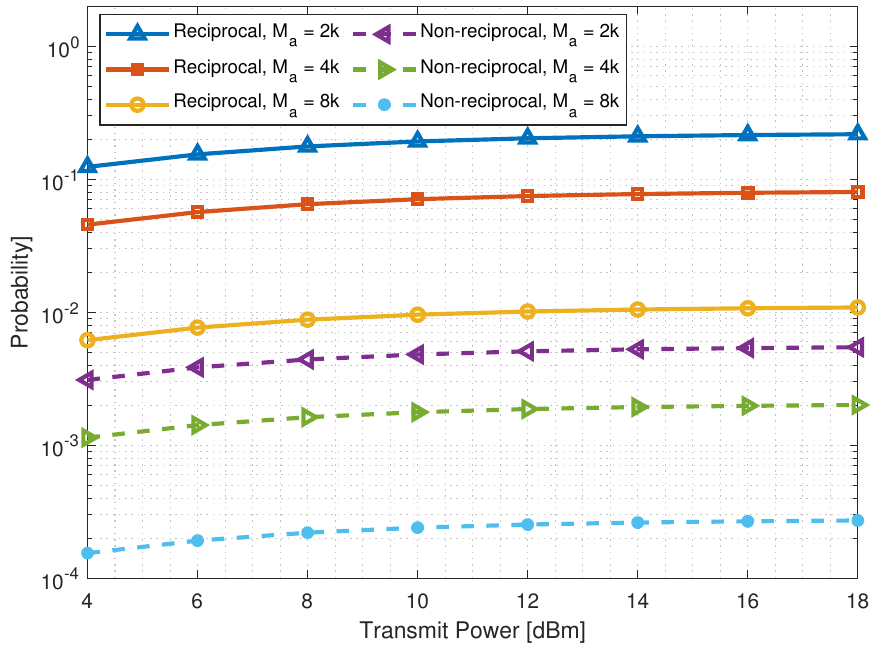}{1}{fig:resManSiz}{Comparison of the probability of false data detection for different values of $M_{a}$ and $M_{e} = 2000$.}

\section{Conclusions}
\label{sec:conclusion}

A novel defensive system against RITM attacks relying on a non-reciprocal channel via D-RIS is detailed in this work to provide a more reliable and secure communication system. The proposed technique is not only able to prevent eavesdropping, but it also is capable of checking the integrity of injected symbols by a potential adversary without assuming any knowledge of them, unlike the existing works. The employed precoding and combining techniques based on the combination of the CSI of a non-reciprocal channel are robust against eavesdropping and false data injection since its disruption is significantly harder than existing techniques. Moreover, a low overhead CEP for this non-reciprocal channel based on phase flipping is also given. The legitimate entities can easily estimate both DL and UL channels by transmitting some reference signals and exchanging the common dynamic phase of the RIS used in the DL and UL. Hence, the CSI feedback process is avoided since it neither exposes critical information nor facilitates the intrusion process.

Consequently, the novel defensive system relying on a non-reciprocal channel via RIS is not only capable of preventing any potential adversary disruption, but it can also take advantage of all the existing benefits provided by the RIS given in the literature. The integration of this defensive system is a key element in the evolution towards 6G since it protects the existing communication system, safeguarding the privacy of users.

% or


\appendices

%\section*{Acknowledgment}
%This work has been partially funded by project TERESA-ADA (TEC2017-90093-C3-2-R) (MINECO/AEI/FEDER, UE), and the work of A. M. Tonello has been supported in part by the Chair of Excellence Program of the Universidad Carlos III de Madrid.

% by themselves when using end float and the captions off option.
\ifCLASSOPTIONcaptionsoff
  \newpage
\fi

\end{document}